\documentclass[journal,10pt]{IEEEtran}
\usepackage{cite}
\usepackage{graphicx}
\usepackage{epstopdf}
\usepackage{amsmath,amssymb,amsfonts}
\usepackage{algorithmic}
\usepackage{textcomp}
\usepackage{xcolor}
\usepackage{tabularx,booktabs}
\usepackage[flushleft]{threeparttable}

\usepackage{booktabs,ragged2e}
\usepackage{multirow}
\usepackage{subfigure}

\usepackage{bm}
\usepackage[colorlinks,
            linkcolor=black,
            anchorcolor=black,
            citecolor=black
            ]{hyperref}

\def\BibTeX{{\rm B\kern-.05em{\sc i\kern-.025em b}\kern-.08em
    T\kern-.1667em\lower.7ex\hbox{E}\kern-.125emX}}

\begin{document}

\title{Evolving Semantic Communication with Generative Model}

\author{
Shunpu Tang,
Qianqian Yang,
Deniz Gündüz,
Zhaoyang Zhang

 \thanks{S. Tang,  Q. Yang, and Z. Zhang are all with the College of Information Science and Electronic Engineering, Zhejiang University, Hangzhou, China (email: \{ qianqianyang20, tangshunpu, zhzy\}@zju.edu.cn).}
 \thanks{ D. Gündüz is with the Department of Electrical and Electronic Engineering, Imperial College London, London, UK, SW7 2AZ. (email: d.gunduz@imperial.ac.uk).}

}

%


\maketitle
\thispagestyle{empty}
\pagestyle{empty}
\begin{abstract}

Recently, learning-based semantic communication (SemCom) has emerged as a promising approach in the upcoming 6G network and researchers have made remarkable efforts in this field. However, existing works have yet to fully explore the advantages of the evolving nature of learning-based systems, where knowledge accumulates during transmission have the potential to enhance system performance. In this paper, we explore an evolving semantic communication system for image transmission, referred to as ESemCom,  with the capability to continuously enhance transmission efficiency.  The system features a novel channel-aware semantic encoder that utilizes a pre-trained Semantic StyleGAN to extract the channel-correlated latent variables consisting of serval semantic vectors from the input images, which can be directly transmitted over a noisy channel without further channel coding. Moreover, we introduce a semantic caching mechanism that dynamically stores the transmitted semantic vectors in the local caching memory of both the transmitter and receiver. The cached semantic vectors are then exploited to eliminate the need to transmit similar codes in subsequent transmission, thus further reducing communication overhead. Simulation results highlight the evolving performance of the proposed system in terms of transmission efficiency, achieving superior perceptual quality with an average bandwidth compression ratio (BCR) of 1/192 for a sequence of 100 testing images compared to DeepJSCC and Inverse JSCC with the same BCR. Code of this paper is available at \url{https://github.com/recusant7/GAN_SeCom}.

\end{abstract}

\begin{IEEEkeywords}
Semantic communication, wireless cache, GAN, joint source-channel coding.
\end{IEEEkeywords}

\section{Introduction}
Recently, there has been a growing interest in semantic communications (SemCom) as a promising technology poised to boost communication efficiency in the forthcoming 6G networks\cite{Semantic1}. In contrast to conventional digital communication systems that treat all transmitted bits uniformly and concentrate solely on bit accuracy, semantic communication prioritizes the transmission and reconstruction of data with more valuable semantic information. This approach aligns well with the increasing demands of machine-to-machine (M2M) communication, which provides significant reductions in transmitted data and is particularly beneficial for intelligent applications such as smart cities and extended reality (XR).


Given the remarkable success of deep learning technology, a semantic communication system for image transmission, named DeepJSCC, was initially investigated in \cite{DeepJSCC}. In DeepJSCC, the joint source-channel encoder and decoder are modeled as neural networks, and trained jointly to optimize the mean square error (MSE) or the structural similarity (SSIM) between the transmitted and reconstructed images. In a related study \cite{zhang2022wireless}, the authors proposed transmitting high-level semantic information, such as text descriptions, to enhance image transmission performance, especially in bandwidth-limited scenarios. DeepJSCC was further extended to the digital communication system in \cite{DeepJSCC-Q}, enhancing its compatibility for deployment in existing communication systems.

A promising approach for boosting the transmission efficiency is to harness the capabilities of recently emerged deep generative models, capable of producing high-dimensional realistic multimedia information from low-dimensional data. 
The authors in \cite{han2022semantic} utilize a deep speech generation model to reconstruct speech from low-dimensional semantic information received, thereby decreasing the required bandwidth to only 0.2\% of that achieved by existing works. The approach presented in \cite{tianxiao_GAN} employs the generative adversarial network (GAN) inversion method \cite{Gan_inversion} to acquire a low-dimensional latent representation of images. Subsequently, the image is reconstructed at the receiver using a GAN, achieving perceptual quality comparable to existing works while utilizing only a small fraction of the channel bandwidth. Works in \cite{GAN_JSCC, yilmaz2023high, chen2023commin} employ a pretrained generative model at the receiver to ensure that the receiver generates a realistic reconstruction of the source images. 

While the aforementioned approaches have achieved remarkable results, SemCom transceivers are typically trained offline and become fixed after deployment in real-world networks. Consequently, the performance of these systems often remains stable or may even decline when confronted with a mismatch between the source or channel characteristics and the training data. One key advantage of data-driven approaches is their ability to continuously improve and adapt to a changing environment with accumulated data. Therefore, we are motivated to design a SemCom approach that can evolve based on the accumulated transmission experience, aiming to further reduce the transmission overhead.

In particular, we integrate wireless cache technology into the SemCom system to store previously transmitted semantic information from the transmitter to the receiver. This stored information is later leveraged to avoid transmitting similar semantic data, resulting in a reduction in communication overhead. While wireless caching has been extensively studied in digital communication systems, such as \cite{cache_1, cache_2}, its application in SemCom systems is a novel aspect of the proposed approach. \textcolor{black}{Moreover, we can extract disentangled semantic information using existing GAN method, which enables us making cache decisions for the disentangled semantic information in a more refined way than for the whole source signals, making it easier to achieve potential caching gains. }

In this paper, we investigate a cache-enabled ESemCom system leveraging the pre-trained semantic StyleGAN \cite{SemanticStyleGAN} model, where both the transmitter and the receiver are equipped with cache memories. Unlike previous works\cite{GAN_JSCC,Gan_inversion}, we introduce a novel channel-aware GAN inversion method to extract the meaningful semantic information from the original input and map it into the channel-correlated latent space. This approach eliminates the need for additional channel encoding and decoding, thereby avoiding additional training overhead related to various channel signal-to-noise ratios (SNR). We also propose a semantic caching mechanism that utilizes the cache memory at both the transmitter and receiver, which stores disentangled semantic vectors at both ends, following the first-in-first-out (FIFO) principle. When a similar semantic vector exists in the caching memory, the transmitter sends the corresponding index instead, thereby reducing communication overhead. To validate our approach, we conduct simulation experiments focusing on the transmission of face images using the CelebA-Masked dataset.  Simulation results demonstrate the effectiveness of the proposed system in terms of peak signal-to-noise ratio (PSNR), perceptual quality and transmission efficiency. Furthermore, the results show a continuous improvement in the transmission efficiency of our proposed method, facilitated by the introduced caching mechanism.

\section{Problem Statement}
\subsection{System Model}
In this paper, we consider the problem of transmitting images over a noisy channel, where both the transmitter and receiver are equipped with a cache memory of same size. The original image can be represented as $\bm{x} \in \mathbb{R}^{3 \times N_H \times N_W}$, where $3$ represents RGB channels, while $N_H$ and $N_W$ are the height and width of the image, respectively. The transmitter uses a GAN-based semantic encoder to extract the latent variable $\bm{z} \in \mathbb{R}^{N_S \times N_L}$ from $\bm{x}$ through the GAN inversion method\cite{Gan_inversion}, which can be expressed as
\begin{equation}
    \bm{z}=G^{-1}(\bm{x}),
\end{equation}
where there are $N_S$ distinct semantic vectors in $\bm{z}$, and $N_L$ is the length for each semantic vector. \textcolor{black}{For example, different semantic vectors in the latent variable of a facial image can represent different semantic information such as eyes, eyebrows, mouth, nose, etc}. Notation $G^{-1}(\cdot)$ denotes the inversion process of the generative function $G(\cdot)$. After that, those semantic vectors that are close enough to one of the cached semantic information in the caching memory are eliminated to reduce the communication overhead. We refer to this process as the semantic caching operation, given by
\begin{equation}
    \tilde{\bm{z}}=C(\bm{z}),
\end{equation}
where $C(\cdot)$ represent the semantic caching operation, $\tilde{\bm{z}}\in \mathbb{R}^{n_s \times N_L}$, and  $n_s\leq N_S$ depends on how many among the $N_S$ semantic vectors are eliminated. \textcolor{black}{We assume that the cache memory at both the transmitter and the receiver can store $N_C$ realizations of each semantic vector.}

To meet the average transmit power constraint, power normalization (PN) \cite{DeepJSCC} is applied to the latent variable $\tilde{\bm{z}}$, before being mapped to the complex-valued channel input signals $\bm{z}_c \in \mathbb{C}^{k}$ for transmission. We have $k = \frac{1}{2} \times n_s \times N_L$, where $\frac{1}{2}$ comes from combining two real values to form a single complex channel symbol. We define the bandwidth compression ratio (BCR) as $k/(3 \times N_H \times N_W)$.

We consider an additive white Gaussian noise (AWGN) channel, where the received signal at the receiver is given by
\begin{equation}
    \bm{\hat{z}}_c=\bm{z}_c+\bm{n},
\end{equation}
where $\bm{n}$ is the additive channel noise, consisting of independent complex Gaussian components $\mathcal{CN}(0, \sigma^2)$. The inverse semantic caching operation is applied to recover the dimension of the original latent variable, given by 
\begin{equation}
    \hat{\bm{z}}=C^{-1}(\bm{\hat{z}}_c),
\end{equation}
where $\hat{\bm{z}} \in \mathbb{R}^{N_S \times N_L}$. Then, the pre-trained GAN acts as the semantic decoder to reconstruct the image $\hat{\bm{x}}$ using the generative function, given by
\begin{equation}
    \hat{\bm{x}}=G(\hat{\bm{z}}).
    \label{eq::G}
\end{equation}
\subsection{Metrics}
We evaluate the quality of the reconstructed image $\hat{\mathbf{x}}$ using PSNR, which can be written as
\begin{equation}
    \text{PSNR}(\bm{x},\hat{\bm{x}}) = 10 \log_{10} \left( \frac{{255^2}}{{\text{MSE}(\bm{x},\hat{\bm{x}})}} \right),
\end{equation}
where 255 is the maximum possible pixel value in the image and MSE$(\cdot,\cdot)$ is the mean squared error between the original and reconstructed images. 
We also use the learned perceptual image patch similarity (LPIPS)\cite{LPIPS} to assess the reconstruction performance, which is a metric that has been shown to closely capture the  human perceptual similarity between two images through a pre-trained VGG network, given by
\begin{equation}
     \text{LPIPS}(\mathbf{x},\hat{\mathbf{x}}) = \sum_{l}\frac{1}{N_{H_l} N_{W_l}} \sum_{h,w}^{N_{H_l},N_{W_l}}  || \omega_l \odot (y_{h,w}^{l}-\hat{y}_{h,w}^{l})||^2_2
\end{equation}
where $y_{h,w}^{l}$ and $\hat{y}_{h,w}^{l}$ are the $(h,w)$-th element in the output feature map of the $l$-th VGG layer for $\bm{x}$ and $\hat{\bm{x}}$, respectively. Besides, $N_{H_l}$ and $N_{W_l}$ denote the height and width of the output feature map, and $\omega_l>0$ is the weight to control the importance for the outputs of different layers. Given these two metrics, this paper aims to design semantic encoding and decoding together with an efficient semantic caching mechanism to minimize the required BCR to achieve a desired average image reconstruction quality.

\section{Proposed Method}
\begin{figure*}[!t]
    \centering
    \includegraphics[width=0.85\linewidth]{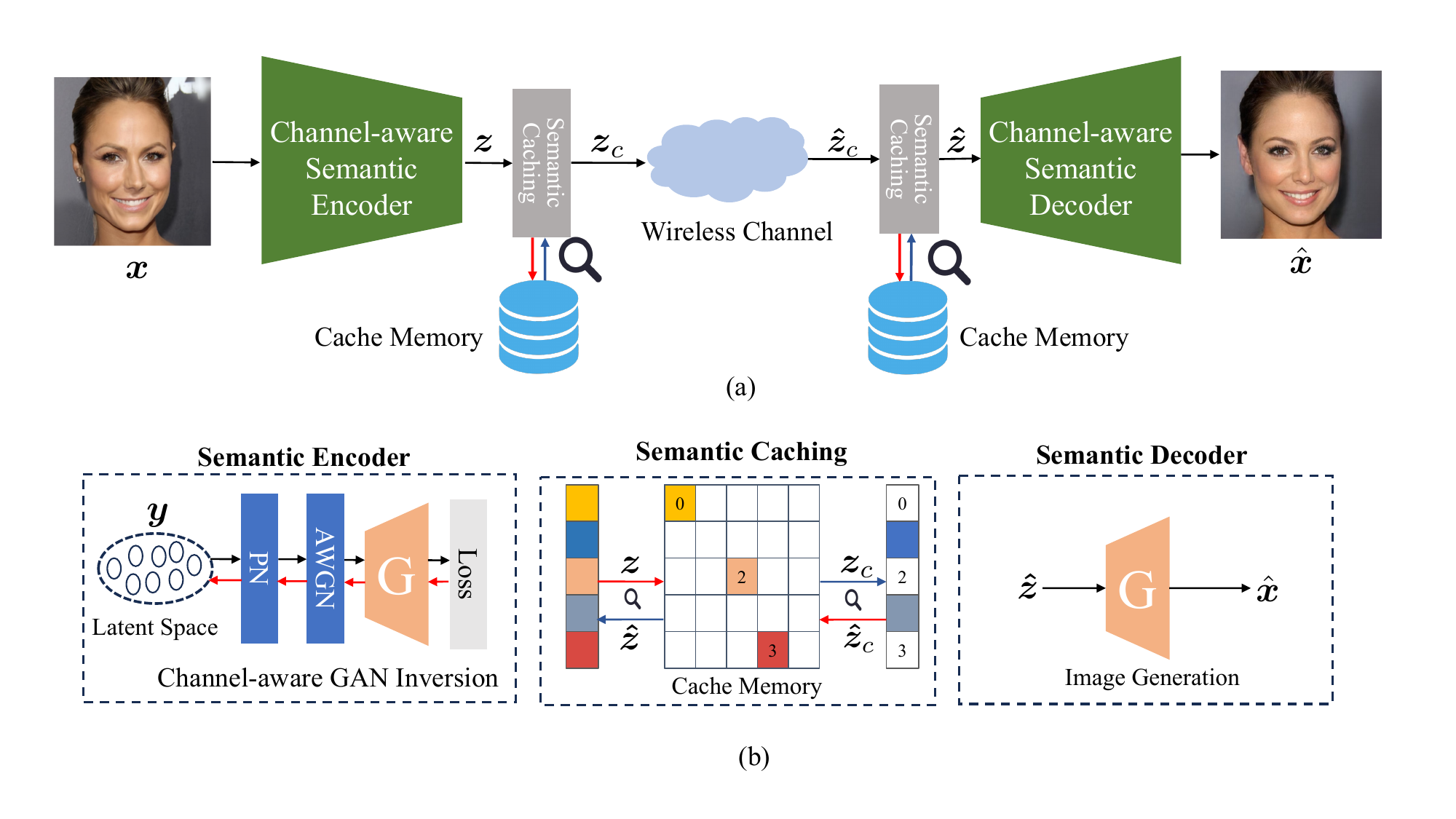}
    \caption{Illustration of the proposed cache-enabled evolving SemCom system}
    \label{fig:overview}
\end{figure*}
In this section, we present the details of the proposed method, including both the channel-aware GAN-based semantic encoder and decoder, as well as the proposed semantic caching mechanism. 

\subsection{Channel-aware Semantic Encoder and Decoder}
For the semantic encoder, we exploit the GAN inversion method, which is the inversion operation of a generative function, to extract semantic information from the given data sample. Specifically, for a given generator denoted by $G$, which maps a latent variable to a data sample, the inversion operation seeks the latent variable $\bm{z}$ of a data sample $\bm{x}$ by 
\begin{equation}
    \bm{z}=\arg\min_{\bm{y}} ||G(\bm{y}) - \bm{x}||^2_2.
    \label{eq::inversion_1}
\end{equation}
In our SemCom system, the derived $\bm{z}$ can be considered as important semantic information, given its significantly lower dimensionality compared to the original sample $\bm{x}$. Building upon our prior research \cite{tianxiao_GAN}, we utilize the semantic StyleGAN2\cite{SemanticStyleGAN} as the backbone generative model, which enables us to extract the disentangled latent variable, with each semantic vector representing a distinct semantic feature.


However, $\bm{z}$ cannot be directly transmitted over the channel as it may not meet the average transmit power constraint. Moreover, when it gets corrupted by the channel noise, the reconstruction performance may deteriorate significantly. Therefore, we introduce a novel channel-aware GAN inversion method, which incorporates the characteristics of the AWGN channel into the GAN inversion process. Compared with the previous work \cite{tianxiao_GAN}, which applies a channel encoder with normalizing flow to encode $\bm{z}$, the proposed method eliminates the need for training additional channel encoder and decoder. This both reduces the training requirements and expedites the end-to-end transmission process. Additionally, the transmitter and receiver only require a pre-trained generator to adapt to various channel conditions, without the need for any re-training or fine-tuning operations.

The process of channel-aware semantic coding is illustrated in \autoref{fig:overview}. We first sample a latent variable $\bm{y}$ from the latent space. Instead of directly feeding it into $G$, we incorporate the PN operation to control its average power. Next, AWGN $\bm{n}$ is added to $z$ to simulate corruption from the noisy channel. We note that here it is assumed that the transmitter has the knowledge of the current noise power in AWGN channel. Hence, for a given image $\bm{x}$, we want to find the latent variable  $\bm{z}$ by solving the following optimization problem, 
\begin{equation}
    \bm{z}= \arg\min_{\bm{y}} ||G ({PN(\bm{y})+\bm{n}}) - \bm{x}||^2_2.
    \label{eq::inversion_2}
\end{equation}
To preserve more perceptual quality, we replace the L2 loss in \eqref{eq::inversion_2} with L1 loss and incorporate the LPIPS loss term, resulting in
\begin{equation}
\begin{aligned}
        \bm{z}=&\arg\min_{\bm{y}} \lambda_1 ||G ({PN(\bm{y})+\bm{n}}) - \bm{x}||\\
        & +\lambda_2 \text{LPIPS}( G ({PN(\bm{y})+\bm{n}}),\bm{x}),
    \label{eq::inversion_3}
\end{aligned}
\end{equation}
where $\lambda_1$ and $\lambda_2$ are positive hyper-parameters controlling the trade-off between the two losses. Compared to the MSE loss function, the L1 loss and LPIPS loss used here align more closely with human perceptual characteristics. We can optimize \eqref{eq::inversion_3} using the gradient descent method, thus deriving $\bm{z}$ for transmission over the noisy channel. 

\textcolor{black}{The semantic decoder at the receiver is simply the corresponding generator $G$ to the GAN inversion $G^{-1}$ used at the transmitter. After the semantic caching operation at the transmitter, which will be explained in more detail in the next section, the noisy signal $\hat{z}$ is directly input into the semantic decoder to generate the original image, without any additional denoising operation.}
\subsection{Semantic Caching Mechanism}
We store a set of semantic vectors from previous transmissions in the cache memory of both the transmitter and the receiver. Upon obtaining $\bm{z}$, the transmitter searches within the cached semantic vectors and replaces one or more components in the latent variables with the corresponding indices if similar semantic vectors exist in the local cache memory. This further reduces communication overhead.

In particular, let $\bm{z}_i\in \mathbb{R}^{N_L}$,  $0\leq i < N_S$, denote the $i$-th semantic vector of latent variables $\bm{z}$. For each vector, $\bm{z}_i$, the transmitter searches in its local cache memory for the closest cached semantic vector, evaluated by cosine similarity, and obtains its index $j^*$ as follows
\begin{equation}
\begin{aligned}
    j^* =& \arg\max_{0\leq j < N_C } \mathrm{cosine}(\bm{z}_i, \bm{c}_{i,j}) \\
    = &\arg\max_{0\leq j < N_C } \frac{\bm{z}_i\cdot \bm{c}_{i,j}}{|\bm{z}_i||\bm{c}_{i,j}|}.
\end{aligned}
\end{equation}
If the cosine similarity $\mathrm{cosine}(\bm{z}_i, \bm{c}_{i,j^*})$ is greater than a pre-defined threshold $\gamma_{i}$, we use the index $j^*$ to replace the original $\bm{z}_i$ for transmission. This operation is executed for all dimensions, resulting in $\tilde{\bm{z}}\in \mathbb{R}^{n_{s}\times N_L}$, where
\begin{equation}
     n_s=\sum_{0 \leq i < N_S} \mathbb{I} \bigg(  cosine(\bm{z}_i, \bm{c}_{i,j^*})\geq \gamma_{i}\bigg),
\end{equation}
and $\mathbb{I}(\cdot)$ denotes the indicator function. {\color{black}The transmitter then sends $\tilde{\bm{z}}$, together with the corresponding indices. We code each index with $\log_2{N_S}+\log_2{N_C}$ bits, which is  transmitted using digital communication technique.} For those components in $\bm{z}$ for which we do not find a replacement in the cache memory, the transmitter stores them in the cache memory. For simplicity, we employ the widely used first-in-first-out (FIFO) principle to update the cache memory with the new latent variables.

Upon receiving $\bm{\hat{z}}_c$ and the indices, the receiver first transforms the complex-value latent variable back to the real-value one. Then, the receiver identifies the corresponding semantic vector using the received indices, thus obtaining the complete latent variable $\hat{\bm{z}}$. The receiver also updates the cache memory with the semantic vector that have been received, ensuring synchronization between the cache contents at the transmitter and the receiver. {\color{black}We note that in this way the stored semantic vectors in the cache memories of the transmitter and receiver would be different due to the impact of channel noise. However, the simulation results reveal that the proposed caching mechanism still helps in terms of transmission efficiency.}. 
 \begin{figure}[t!]
    \centering
    \includegraphics[width=0.8\linewidth]{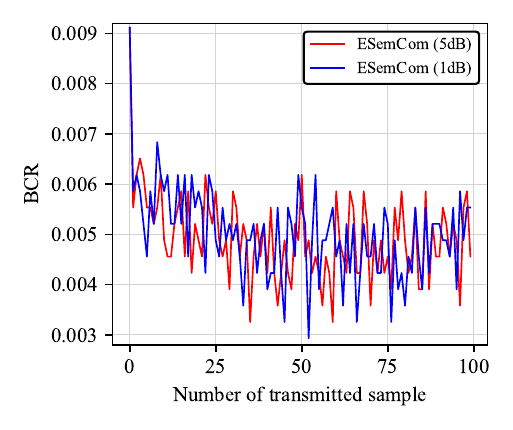}
    \caption{BCR versus the index of the transmitted images, tested SNR=5dB.}
    \label{fig:BCR}
\end{figure}
\begin{figure}[t!]
    \centering
     \includegraphics[width=0.8\linewidth]{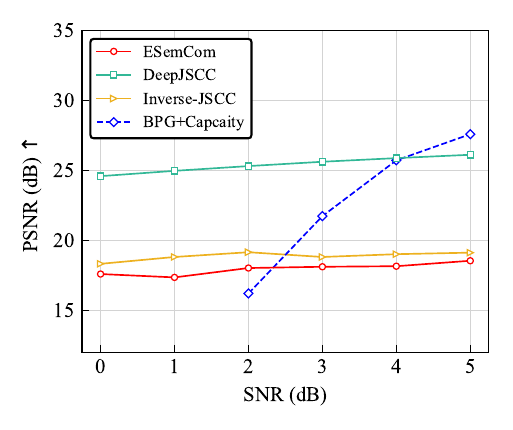}
     \caption{Image reconstruction quality evaluated by PSNR versus SNR.}
      \label{fig:PSNR}
\end{figure}

\begin{figure}[t!]
    \centering
     \includegraphics[width=0.8\linewidth]{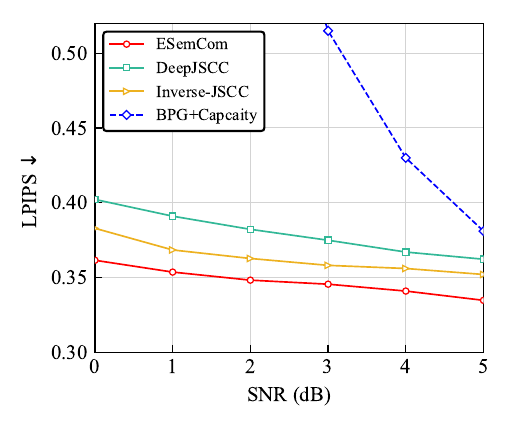}
     \caption{Image reconstruction quality evaluated by LPIPS versus SNR.}
      \label{fig:LPIPS}
\end{figure}


\begin{figure*}[ht!]
    \centering
    \includegraphics[width=\textwidth]{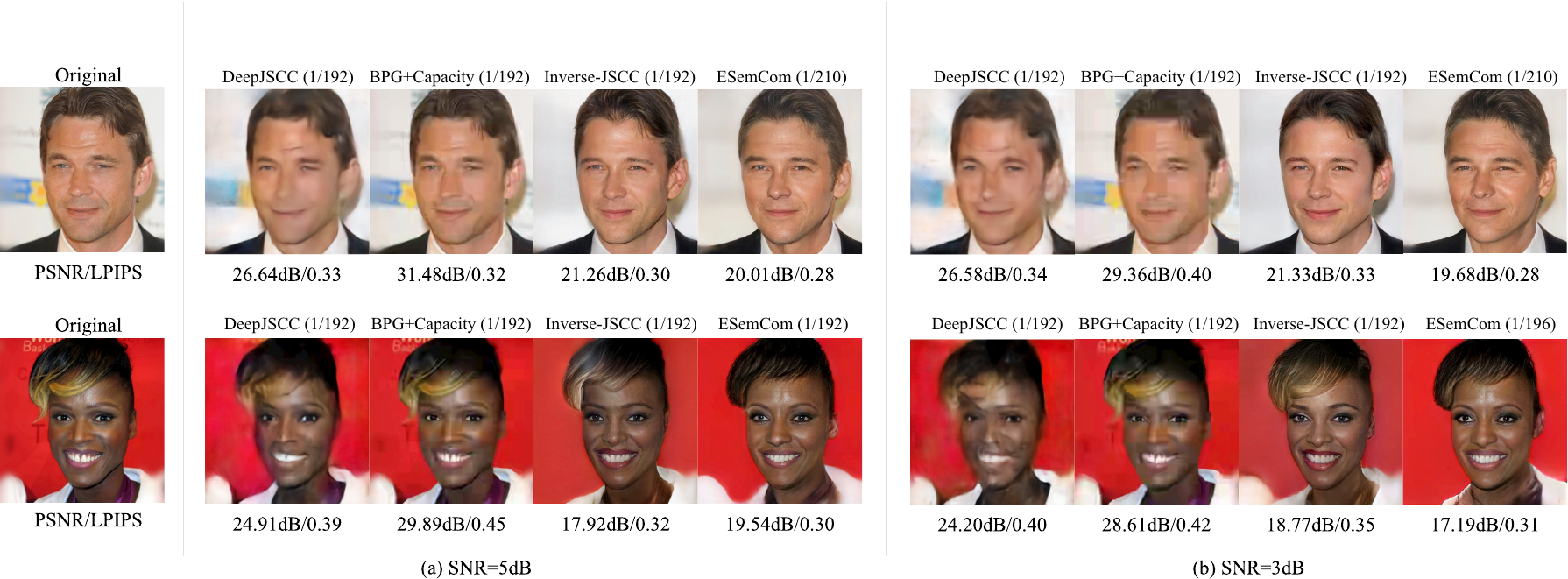}
    \caption{
Visual comparison on the reconstructed CelebAMask-HQ images under SNR of 5dB and 3dB. }
    \label{fig:faces}
\end{figure*}

\section{Simulation}
In this section, we conduct simulations to illustrate the effectiveness of the proposed method. Firstly, we outline the simulation settings, which include details about the dataset used and the key parameters of the transmission model. We then assess the performance of the proposed method in comparison to the existing SemCom system.
\subsection{Simulation Settings}
In the simulation, we set the transmit power to unity, while allowing the SNR of the AWGN channel to vary between -5dB and 5dB. Both the transmitter and receiver are equipped with a cache memory capable with $N_C=50$, and hence can store 50 realizations of each semantic vector in latent variable $\bm{z}$. Additionally, we assume that the transmitter has accurate knowledge of the current channel noise power. 

We employ semantic styleGAN as the backbone generative model, which has been trained on CelebAMask-HQ\cite{CelebAMask-HQ} dataset, consisting of 30,000 high-resolution facial images, each of size $3\times512\times512$. The dimension of the latent variables $\bm{z}$ is set to $N_S= 28$ and $N_L=512$. The similarity threshold $\gamma_{i}$, is set to 0.95 for $i=6, ..., 13$, which represents more semantically important facial regions such as eyes, eyebrows, mouth, and nose. For the remaining $\gamma_{i}$ values, we set them to 0.8. 
We note that the proposed model achieves a BCR of $1/110$ if $\bm{z}$ is directly transmitted after real-to-complex value transformation mentioned earlier. This BCR value can be further reduced with the proposed semantic caching mechanism.

We conduct simulations for the first 100 images from the testing dataset of CelebAMask-HQ. Notably, these images are not used to pre-train the semantic styleGAN. We compare the LPIPS performance between the proposed ESemCom with the following methods,
\begin{itemize}
    \item \textbf{DeepJSCC}\cite{DeepJSCC}: The classical SemCom method for image transmission.
    \item  \textbf{Inverse-JSCC}\cite{GAN_JSCC}: Extended version of DeepJSCC, which introduces the GAN inversion method into SemCom system. We use the same semantic \textit{styleGAN} as the generative model in this method, and follow \cite{GAN_JSCC} to train the semantic encoder and decoder using the loss function of $||\bm{x}-\hat{\bm{x}}||^2_2+\text{LPIPS}(\bm{x},\hat{\bm{x}})$. 
    \item \textbf{BPG+Capcaity:} The upper bound performance on any digital communication system that use the SOTA image compression algorithm named better portable graphics (BPG)\footnote{\url{https://bellard.org/bpg/}} for source coding, which compress images at the rate of channel capacity and assumes error-free transmission in this case.
\end{itemize}
\textcolor{black}{For a fair comparison, we take into account the bandwidth cost for transmitting the indices by the proposed method. In particular, we use channel coding with a rate of 1/2 and binary phase-shift keying (BPSK), which can achieve a successful transmission rate of $p=0.9$ per block under an SNR of 0dB when the block length is more than 128 according to \cite{Short_Block_Length_Codes}. The number of transmissions for successfully sending  indices follows a geometric distribution with probability $p$. We then have the expected number of transmissions to be $1/p$. Therefore, the channel use for each index is $\frac{1}{p}\times 2\times (\log_2{N_S}+\log_2{N_C})$.} 




\subsection{Simulation Results}




We transmit the testing 100 images one by one under the AWGN channel with a SNR of 5dB and 1dB, respectively. As shown in \autoref{fig:BCR},  for the initial transmission with empty cache memories, the BCR for the first image by the proposed ESemCom scheme is approximately $1/110$. As the cache memories at the transmitter and receiver are populated with semantic vectors, the BCR by ESemCom quickly decreases. Notably, the smallest BCR reaches $1/260$, and the average BCR of transmitting these 100 images is about 1/192. This can be attributed to the proposed semantic caching mechanism, by which transmitted latent variables are cached and replaced with similar ones in subsequent transmissions, thus avoiding redundant transmissions and improving communication efficiency. \textcolor{black}{We note that on average, 15.3 semantic vectors in $\bm{z}$ are replaced by indices, requiring only 370 symbols on average for transmitting these indices.} We observe fluctuations in the BCR values of ESemCom, which arise from certain images exhibiting more similarities in their semantic vectors with respect to the images transmitted earlier. These results demonstrate the evolving characteristics of ESemCom and emphasize the effectiveness of the proposed semantic caching mechanism.

We then compare the performance of the proposed ESemCom, DeepJSCC, Inverse-JSCC, and BPG+Capacity in terms of image reconstruction quality, evaluated by PSNR and LPIPS, as shown in \autoref{fig:PSNR} and \autoref{fig:LPIPS}, where an AWGN channel with SNRs ranging from 0 to 5dB is considered during testing. DeepJSCC and Inverse-JSCC are trained with uniform SNRs from 0 to 5dB. The average BCR of ESemCom is about 1/192, and hence we set the BCR of the other methods to be 1/192 for a fair comparison. From these figures, the proposed ESemCom and Inverse-JSCC achieve a lower PSNR than DeepJSCC, but outperform DeepJSCCC and BPG+Capcailty in terms of LPIPS. This is because these methods emphasize prioritizing semantic information while disregarding irrelevant or less critical image details. Moreover, the proposed ESemCom achieves better LPIPS than Invese-JSCC, and still comparable LPIPS when SNR is 0dB. These results further validate that the proposed system effectively transmits images while preserving perceptual quality, especially in challenging channel conditions.


We further provide a visual comparison on the reconstructed CelebAMask-HQ images under test SNR of 5dB and 3db, respectively, in \autoref{fig:faces}. From this figure, we can find that the proposed ESemCom and Inversion-JSCC exhibit superior human perceptual quality with the help of the generative model. Specifically, the proposed scheme can persevere intricate facial details, including features such as eyes, nose, and mouth, while effectively filtering out irrelevant background information. In contrast, facial regions generated by DeepJSCC appear blurred and even unrecognizable under this challenging condition of low BCR and SNR. Moreover, lots of artifacts occur in the images generated BPG+Capacity. These results further demonstrate the superiority of the proposed ESemCom scheme.
\begin{figure}[t!]
    \centering
    \includegraphics[width=\linewidth]{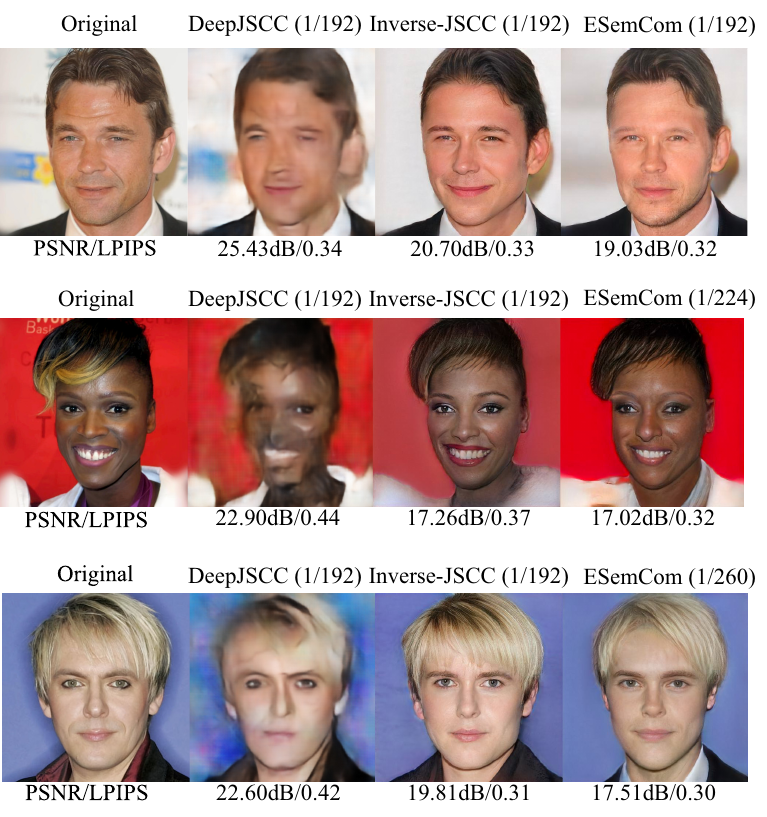}
    \caption{Visual comparison on the reconstructed CelebAMask-HQ images under SNR of 0dB. }
    \label{fig:face2}
\end{figure}

\textcolor{black}{\autoref{fig:face2} illustrates the visual comparison at a testing SNR of 0dB. It is evident from this figure that the proposed ESemCom and Inverse-JSCC still achieve comparable image quality in this challenging channel condition, while DeepJSCC fails to perform well and BPG+Capacity is unable to transmit images. These results demonstrate the robustness of generative model-based methods in challenging channel conditions.}

\section{Conclusion}

In this paper, we explored an evolving SemCom system designed to continuously improve its transmission efficiency by utilizing cache memory at both the transmitter and receiver and a pretrained generative model. In particular, we introduced a novel channel-aware semantic encoder, employing a GAN inversion method to map input images into a channel-correlated space. This approach allows direct transmission of the obtained latent variables over a noisy channel without additional channel encoding, substantially improving system robustness under diverse channel conditions. Additionally, we proposed a semantic caching mechanism, improving communication efficiency by avoiding the transmission of semantic vectors already present in the local cache memory. Simulation results demonstrate the superiority of our proposed system, showing improved perceptual quality with approximately half the BCR compared to DeepJSCC.

\bibliographystyle{IEEEtran}
\bibliography{IEEEabrv, references}

\begin{thebibliography}{10}
\providecommand{\url}[1]{#1}
\csname url@samestyle\endcsname
\providecommand{\newblock}{\relax}
\providecommand{\bibinfo}[2]{#2}
\providecommand{\BIBentrySTDinterwordspacing}{\spaceskip=0pt\relax}
\providecommand{\BIBentryALTinterwordstretchfactor}{4}
\providecommand{\BIBentryALTinterwordspacing}{\spaceskip=\fontdimen2\font plus
\BIBentryALTinterwordstretchfactor\fontdimen3\font minus
  \fontdimen4\font\relax}
\providecommand{\BIBforeignlanguage}[2]{{%
\expandafter\ifx\csname l@#1\endcsname\relax
\typeout{** WARNING: IEEEtran.bst: No hyphenation pattern has been}%
\typeout{** loaded for the language `#1'. Using the pattern for}%
\typeout{** the default language instead.}%
\else
\language=\csname l@#1\endcsname
\fi
#2}}
\providecommand{\BIBdecl}{\relax}
\BIBdecl

\bibitem{Semantic1}
D.~G{\"{u}}nd{\"{u}}z, Z.~Qin, I.~E. Aguerri, H.~S. Dhillon, Z.~Yang, A.~Yener,
  K.~Wong, and C.~Chae, ``Beyond transmitting bits: Context, semantics, and
  task-oriented communications,'' \emph{{IEEE} J. Sel. Areas Commun.}, vol.~41,
  no.~1, pp. 5--41, 2023.

\bibitem{DeepJSCC}
E.~Bourtsoulatze, D.~B. Kurka, and D.~G{\"{u}}nd{\"{u}}z, ``Deep joint
  source-channel coding for wireless image transmission,'' \emph{{IEEE} Trans.
  Cogn. Commun. Netw.}, vol.~5, no.~3, pp. 567--579, 2019.

\bibitem{zhang2022wireless}
Z.~Zhang, Q.~Yang, S.~He, M.~Sun, and J.~Chen, ``Wireless transmission of
  images with the assistance of multi-level semantic information,'' in
  \emph{IEEE Int. Symp. Wire. Commun. Sys. (ISWCS)}, 2022, pp. 1--6.

\bibitem{DeepJSCC-Q}
T.-Y. Tung, D.~B. Kurka, M.~Jankowski, and D.~Gündüz, ``Deep{JSCC-Q}:
  Constellation constrained deep joint source-channel coding,'' \emph{{IEEE} J.
  Sel. Areas Inf. Theory}, vol.~3, no.~4, pp. 720--731, 2022.

\bibitem{han2022semantic}
T.~Han, Q.~Yang, Z.~Shi, S.~He, and Z.~Zhang, ``Semantic-preserved
  communication system for highly efficient speech transmission,'' \emph{{IEEE}
  J. Sel. Areas Commun.}, vol.~41, no.~1, pp. 245--259, 2022.

\bibitem{tianxiao_GAN}
T.~Han, J.~Tang, Q.~Yang, Y.~Duan, Z.~Zhang, and Z.~Shi, ``Generative model
  based highly efficient semantic communication approach for image
  transmission,'' in \emph{IEEE Int. Conf. Acoust. Speech Signal Process.
  (ICASSP)}, 2023, pp. 1--5.

\bibitem{Gan_inversion}
W.~Xia, Y.~Zhang, Y.~Yang, J.-H. Xue, B.~Zhou, and M.-H. Yang, ``{GAN}
  inversion: A survey,'' \emph{{IEEE} Trans. Pattern Anal. Machine Intell.},
  vol.~45, no.~3, pp. 3121--3138, 2022.

\bibitem{GAN_JSCC}
E.~Erdemir, T.-Y. Tung, P.~L. Dragotti, and D.~Gündüz, ``Generative joint
  source-channel coding for semantic image transmission,'' \emph{{IEEE} J. Sel.
  Areas Commun.}, vol.~41, no.~8, pp. 2645--2657, 2023.

\bibitem{yilmaz2023high}
S.~F. Yilmaz, X.~Niu, B.~Bai, W.~Han, L.~Deng, and D.~Gunduz, ``High perceptual
  quality wireless image delivery with denoising diffusion models,''
  \emph{arXiv:2309.15889}, 2023.

\bibitem{chen2023commin}
J.~Chen, D.~You, D.~G{\"u}nd{\"u}z, and P.~L. Dragotti, ``Commin: Semantic
  image communications as an inverse problem with inn-guided diffusion
  models,'' \emph{arXiv:2310.01130}, 2023.

\bibitem{cache_1}
Q.~Yang and D.~G{\"{u}}nd{\"{u}}z, ``Coded caching and content delivery with
  heterogeneous distortion requirements,'' \emph{{IEEE} Trans. Inf. Theory},
  vol.~64, no.~6, pp. 4347--4364, 2018.

\bibitem{cache_2}
S.~Tang, K.~He, L.~Chen, L.~Fan, X.~Lei, and R.~Q. Hu, ``Collaborative
  cache-aided relaying networks: Performance evaluation and system
  optimization,'' \emph{{IEEE} J. Sel. Areas Commun.}, vol.~41, no.~3, pp.
  706--719, 2023.

\bibitem{SemanticStyleGAN}
Y.~Shi, X.~Yang, Y.~Wan, and X.~Shen, ``{SemanticStyleGAN}: Learning
  compositional generative priors for controllable image synthesis and
  editing,'' in \emph{{IEEE/CVF} Int. Comput. Vis. Pattern Recognit. (CVPR)},
  2022, pp. 11\,244--11\,254.

\bibitem{LPIPS}
R.~Zhang, P.~Isola, A.~A. Efros, E.~Shechtman, and O.~Wang, ``The unreasonable
  effectiveness of deep features as a perceptual metric,'' in \emph{{IEEE/CVF}
  Int. Comput. Vis. Pattern Recognit. (CVPR)}, 2018, pp. 586--595.

\bibitem{CelebAMask-HQ}
C.-H. Lee, Z.~Liu, L.~Wu, and P.~Luo, ``Maskgan: Towards diverse and
  interactive facial image manipulation,'' in \emph{{IEEE/CVF} Int. Comput.
  Vis. Pattern Recognit. (CVPR)}, 2020, pp. 5548--5557.

\bibitem{Short_Block_Length_Codes}
M.~Shirvanimoghaddam, M.~S. Mohammadi, R.~Abbas, A.~Minja, C.~Yue, B.~Matuz,
  G.~Han, Z.~Lin, W.~Liu, Y.~Li, S.~Johnson, and B.~Vucetic, ``Short
  block-length codes for ultra-reliable low latency communications,''
  \emph{{IEEE} Commun. Mag.}, vol.~57, no.~2, pp. 130--137, 2019.

\end{thebibliography}

\end{document}